\documentclass[12pt,preprint]{aastex}

\begin{document}
\title {Reappraising the Spite Lithium Plateau: Extremely Thin
and Marginally Consistent with WMAP
} 

\newcommand{\teff}{T$_{\rm eff}$ }
\newcommand{\tirfm}{T$_{\rm eff}^{\rm IRFM}$}
\newcommand{\tfe}{T$_{\rm eff}^{\rm FeI}$}

\newbox\grsign \setbox\grsign=\hbox{$>$} \newdimen\grdimen \grdimen=\ht\grsign
\newbox\simlessbox \newbox\simgreatbox
\setbox\simgreatbox=\hbox{\raise.5ex\hbox{$>$}\llap
     {\lower.5ex\hbox{$\sim$}}}\ht1=\grdimen\dp1=0pt
\setbox\simlessbox=\hbox{\raise.5ex\hbox{$<$}\llap
     {\lower.5ex\hbox{$\sim$}}}\ht2=\grdimen\dp2=0pt\def\simgreat{\mathrel{\copy\simgreatbox}}
\def\simless{\mathrel{\copy\simlessbox}}
\newbox\simppropto
\setbox\simppropto=\hbox{\raise.5ex\hbox{$\sim$}\llap
     {\lower.5ex\hbox{$\propto$}}}\ht2=\grdimen\dp2=0pt
\def\simpropto{\mathrel{\copy\simppropto}}

\author{Jorge Mel\'endez\altaffilmark{1} and Iv\'an Ram\'{\i}rez\altaffilmark{2}}
\altaffiltext{1}{Department of Astronomy, Caltech, M/C 105-24, 
1200 E. California Blvd, Pasadena, CA 91125; jorge@astro.caltech.edu}
\altaffiltext{2}{Department of Astronomy, University of Texas at Austin, RLM 15.306, TX
78712-1083; ivan@astro.as.utexas.edu}

\slugcomment{Submitted to the The Astrophysical Journal}

\begin{abstract}
The lithium abundance in 62 halo dwarfs is determined 
from accurate equivalent widths reported in the literature
and an improved infrared flux method (IRFM) temperature scale. 
The Li abundance of 41 plateau stars (those with \teff $>$ 6000 K) 
is found to be independent of temperature and metallicity, 
with a star-to-star scatter of only 0.06 dex
over a broad range of temperatures (6000 K $<$ \teff $<$ 6800 K) and
metallicities ($-3.4 <$ [Fe/H] $< -1$), thus imposing
stringent constraints on depletion by mixing and
production by Galactic chemical evolution.
We find a mean Li plateau abundance of $A_{Li}$ = 2.37 dex
($^7$Li/H = 2.34 $\times 10^{-10}$), 
which, considering errors of the order of 0.1 dex in the {\it absolute} 
abundance scale, is just in
borderline agreement with the constraints imposed by the 
theory of primordial 
nucleosynthesis and WMAP data (2.51 $< A_{Li}^{\rm WMAP}$ $<$ 2.66 dex).
\end{abstract}


\section{Introduction}
The Li plateau was discovered by Spite \& Spite (1982),
who showed that the $^7$Li abundance obtained from the
Li doublet at 6708 \AA\ in F and early G halo dwarfs is
independent of temperature and metallicity, suggesting
that the Li abundance determined in halo
stars represents the primordial abundance from
Big Bang nucleosynthesis (BBN). The standard
theory of BBN predicts the abundance of light elements,
in particular $^7$Li, as a function of the baryon-to-photon
ratio $\eta$. Thus, if the plateau is primordial, an inverse analysis 
could be used to determine $\eta$ (Spite \& Spite 1982).

On the other hand, Ryan and collaborators (Ryan et al. 1996, 1999, 2001; 
hereafter jointly referred as R01)
suggested that the Li abundance in halo stars is postprimordial, 
with a 0.3 dex increase from [Fe/H] = $-$3.5 to [Fe/H] = $-$1.\footnote{[A/B] 
$\equiv$ log ($N_A/N_B$) - log($N_A/N_B$)$_\odot$}
Those results rely on the 
temperature scale of Magain (1987), which was calibrated with only
3 metal-poor ([Fe/H] $<$ -2) stars.
Considering the metallicity dependence of the Li
abundances, R01 extrapolated a primordial 
Li abundance of $A_{Li}$
$\approx$ 2.0 - 2.1 dex.\footnote{$A_{X}$ $\equiv$ log ($N_X$/$N_H$) + 12}

Recently, precise measurements of the cosmic background 
anisotropies by the Wilkinson Microwave Anisotropy Probe 
(WMAP, Spergel et al. 2003) have been used to constrain $\eta$, 
and employing the standard theory of BBN, the primordial Li abundance 
was predicted to be 
$A_{Li}^{\rm WMAP}$ = 2.58$_{-0.07}^{+0.08}$ dex (Cyburt et al. 2003),
a value that is much higher than the primordial abundance proposed by
R01. The large difference of about
0.5 dex between the results of Ryan et al. and WMAP has
stimulated theoretical work on non-standard BBN
(e.g. Feng et al. 2003; Ichikawa \& Kawasaki 2004).

Bonifacio \& Molaro (1997, hereafter BM97) found that when the \teff scale based
on the infrared flux method (IRFM) is employed (that of Alonso et al. 1996), 
the Li abundance in metal-poor stars does not depend on metallicity, and 
$A_{Li}$ = 2.24 dex is obtained, higher than the primordial
abundance proposed by R01, but still
lower by a factor of 2 than the abundance suggested by the
WMAP data. Furthermore, Bonifacio et al. (2002) analyzed 5 metal-poor field stars
and 12 turn-off stars in the globular cluster NGC 6397,
obtaining a Li abundance 0.05 and 0.1 dex higher than in BM97,
respectively.

Ford et al. (2002) have determined the Li abundance in 
a few metal-poor stars employing the extremely weak 
($\approx$ 2-4 m\AA) Li lines at 6104 \AA, 
resulting in $A_{Li}$ $\approx$ 2.5 dex, but note that the observational
errors are very large (about 30\%), not allowing them to draw
any firm conclusions. In fact, when the Li lines at 6708 \AA\ were used by
Ford et al. (2002), $A_{Li}$ $\approx$ 2.2 dex was found.

The temperature is the most important atmospheric parameter 
required to obtain the Li abundance,
and since metal-poor stars are too distant to measure their
angular diameters, indirect methods have to be used to
estimate their temperatures. The closest (and less model-dependent)
approach to the direct method is the IRFM,
where the temperature is found by a comparison between
the observed and theoretical ratio of
the bolometric flux to the infrared flux. In a pioneer work,
Magain (1987) calibrated the \teff scale of metal-poor
dwarfs (mostly \teff $<$ 6000 K and [Fe/H] $> -2$) employing 
IRFM temperatures of 11 dwarfs. The titanic work by 
Alonso et al. (1996) improved the \teff scale
including a larger number of metal-poor stars 
(down to [Fe/H] $\approx$ -3), employing spectroscopic
metallicities (when available) determined before 1992.
In the meantime, new spectroscopic and photometric
surveys have appeared in the literature 
(a milestone was the completion of 2MASS), 
providing the necessary input data for a new revision of 
the \teff scale (Ram\'{\i}rez \& Mel\'endez 2004b,c, hereafter RM04b,c). 
As an example, only from January 2001 to June 2003, 
more than 1500 new spectroscopic metallicity determinations
for field FGK stars have appeared in the literature, 
increasing significantly the number of calibrating stars with 
reliable [Fe/H] values, and hence allowing to better
define the metallicity dependence of the \teff scale.
We have determined the IRFM \teff of more 
than 10$^3$ stars (RM04b) using updated input data, 
and metallicity-dependent \teff vs. color calibrations 
were derived in the metallicity range $-3.5 <$ [Fe/H] $< +0.4$ (RM04c).

Employing our improved IRFM temperature scale, we show
that the Spite plateau has essentially zero scatter and
that the derived abundances are now consistent, although only
marginally, with WMAP data.

\section{Data and Atmospheric Parameters}
The sample was selected from relatively unevolved halo stars
(log $g \geq$ 3.5, [Fe/H] $<$ $-$1) with
previous Li observations and having either
an IRFM effective temperature determined by us (RM04b)
and/or with a previous determination of metallicity
as reported in the ``2003 updated'' version of the
Cayrel de Strobel et al. (2001, hereafter C03) 
catalog (see $\S$2 in Ram\'{\i}rez \& Mel\'endez 2004a). 

For most works previous to 1997, the equivalent
widths ($W_\lambda$) of the Li doublet at 6708 \AA\ 
were taken from the compilation of Ryan et al. (1996), 
adopting a minimum error of 1.5 m\AA\ for the individual
measurements. 
Other $W_\lambda$ considered are 
those by Molaro et al. (1995), Spite et al. (1996), 
Nissen \& Schuster (1997), Ryan \& Deliyannis (1998), Smith et al. (1998), 
Guti\'errez et al. (1999), Hobbs et al. (1999),
 King (1999),
Ryan et al. (1999, 2001), 
Bonifacio et al. (2002), and Ford et al. (2002).
The weighted averages of the equivalent widths and
errors ($W_\lambda$ and $\delta W_\lambda$) were computed,
as well as the standard errors ($\sigma_m$). The adopted error is
$\sigma_W \equiv$  max($\delta W_\lambda$, $\sigma_m$, 1 m\AA). 
A star was rejected if $\sigma_W \geq 0.1 W_\lambda$.
Stars with E(B-V) $>$ 0.05 dex were also discarded. 
Sixty two dwarfs satisfied our selection criteria.

The key parameter for a reliable Li abundance determination
in metal-poor stars is T$_{\rm eff}$; errors in the other atmospheric parameters 
(log $g$, [Fe/H], $v_t$) result in abundance errors of the order 
of 0.01 dex (e.g. Ryan et al. 1996).
Surface gravities and metallicities were taken
from the C03 catalog, and a microturbulence $v_t$ = 1 km s$^{-1}$ 
was adopted.
For 3 stars not included in the C03 catalog, we
adopted metallicities from a photometric calibration
(RM04b) that is based on C03 [Fe/H] values. 

The effective temperatures are based on our IRFM \teff
scale. Individual IRFM temperatures were
taken from RM04b or recomputed employing new reddening corrections, 
as well as calculated employing the 
IRFM \teff calibrations of RM04c. We used up to 17 colors
in the {\it BVRI}(Johnson-Cousins), Tycho-Hipparcos, Str\"omgren, 
DDO, Vilnius, Geneva and 2MASS systems.
Most of the optical photometry was taken from the General Catalogue of 
Photometric Data (Mermilliod et al. 1997) and
the Tycho catalog, while the infrared photometry was taken from 2MASS.
E(B-V) values were computed employing several extinction maps, 
giving a higher weight to the maps by Schlegel et al. (1998) 
for high latitude stars (RM04b).

The use of several colors improves the determination 
of the effective temperature, since it alleviates the impact of 
photometric errors. Following RM04c, the adopted \teff 
was the mean of the IRFM determination and the 
temperature from the IRFM calibrations (T$_{\rm eff}^{\rm cal}$):

\begin{equation}
{\rm T_{\rm eff}} = \case{1}{n+m} (n {\rm T_{eff}^{IRFM}} + m {\rm T_{eff}^{cal}})
\end{equation}

\noindent where the weights we adopted were $n$ = 1  
and $m$ equal to the square root of the number of colors employed
(typically 10, thus $m \approx 3$).
T$_{\rm eff}^{\rm cal}$ was calculated taking
into account the errors of each IRFM color calibration.
The mean internal error of the \teff derived in
this work is $\delta$\teff $\approx$ 66 K.
It is important to note that the use of eq. (1)
results in a \teff that is only $\approx$ 22 K lower than
the temperature directly determined
from the IRFM. If only T$_{\rm eff}^{\rm IRFM}$ were
used, A$_{Li}$ would be about 0.015 dex higher.

\section{Li Abundances}

In order to determine the Li abundances, we computed a
grid of theoretical equivalent widths of the Li
doublet in the parameters space:
\teff = [5250 K, 5500 K, ..., 6750 K], log $g$ = [3.5, 4.0, 4.5],
[Fe/H] = [-1.0, -1.5, ..., -3.5], $A_{Li}$ = [1.7, 1.9, ..., 2.5].
We took into account the fine and hyperfine structure of the
doublet, with wavelengths and $gf$-values as given in
Andersen et al. (1984). Note that Smith et al. (1998) and
Hobbs et al. (1999) reported new $gf$-values based on
theoretical calculations, but the
difference with our adopted (laboratory) values is only 
$\approx$ 1\% (Smith et al. 1998). The collisional broadening 
constant was obtained from Barklem et al. (2000). 
A grid of 630 synthetic spectra was computed employing the 
2002 version of the LTE spectral synthesis code MOOG (Sneden 1973) 
and Kurucz overshooting models.
The theoretical $W_\lambda$ were measured by integration.

Using the stellar parameters and observed $W_\lambda$
as input, the Li abundance was determined
from a full 4-dimensional interpolation in the synthetic grid
$A_{Li}$(\teff, log $g$, [Fe/H], log $W_\lambda$).
Note that in some previous studies (Thorburn 1994; Ryan et al. 1996, 1999)
a more restricted approach was adopted, 
considering only two fixed log $g$ values (no interpolation) and
metallicity interpolation only for models within $-$2 $<$ [Fe/H] $<$ $-$1 
(Ryan et al. 1996) or $-$3 $<$ [Fe/H] $<$ $-$2 (Thorburn 1994).
Our full interpolation results in lower internal 
abundance uncertainties, 
which is very important to determine 
the true scatter of the Li abundances.

NLTE effects are small for the Li plateau stars.
We applied the NLTE corrections ($\Delta x \equiv$ NLTE - LTE)
by Carlsson et al. (1994), which are
$\Delta x$ $\approx$ $-$0.04 and 0.0 dex for the hottest 
and coolest plateau dwarfs,
respectively. Although these NLTE corrections were not specifically 
calculated for our set of model atmospheres, their relative
dependence with \teff and [Fe/H] is still useful, and
should minimize the scatter of the derived abundances. 
Since the {\it absolute} NLTE corrections may be different, 
we prefer to preserve the mean $A_{Li}$(LTE) of the plateau stars
by adopting a NLTE correction of $\Delta x$ + 0.02 dex.
The final Li abundances are reported in Table 1.

In Fig. 1 are shown the Li abundances obtained in this work.
In the lower panel $A_{Li}$ vs. \teff is plotted, where it is
clearly seen that stars with \teff $>$ 6000 K have approximately
the same Li abundance (plateau stars), as no trend with
\teff (slope of $4 \pm 5 \times 10^{-3}$ per 100 K) is observed. 
In the upper panel are plotted
41 plateau stars (\teff $>$ 6000 K) as a function of [Fe/H], 
showing an impressive constancy in their Li abundance from 
[Fe/H] $\approx$ $-$3.4 to $-$1,
with an amazingly small scatter and
essentially zero slope (0.013 $\pm$ 0.018 per dex).
This figure strongly supports
the primordial origin of the Li plateau (Spite \& Spite 1982),
and imposes stringent constraints on stellar mixing
and Galactic chemical evolution.

The very small star-to-star {\it observed} scatter ($\sigma_{obs}$ = 0.06 dex) 
is fully explained by errors in \teff 
and $W_\lambda$.
Adding in quadrature the typical errors 
of $\delta$T = 66 K ($\delta A_{Li} \approx$ 0.045 dex) 
and $\delta W_\lambda$ = 5.5\% ($\delta A_{Li} \approx$ 0.025 dex)
(as well as errors of 0.3 dex in log $g$ and [Fe/H]), 
a predicted error $\sigma_{pred}$ = 0.053 dex is obtained, 
in excellent agreement with the observed scatter ($\sigma_{obs}$ = 0.06 dex).
Note that if the 2 plateau stars with the highest Li abundances
are discarded, then $\sigma_{obs}$ = 0.05 dex.

\section{Discussion}

The abundances obtained in this work for 
plateau stars show a very small star-to-star-scatter.
This is a consequence of employing very accurate 
temperatures based on the IRFM and their photometric 
calibrations. 
RM04b,c have shown that the IRFM \teff scale is in almost 
perfect agreement (at the 10 K level)
with interferometric measurements of dwarfs and giants with [Fe/H] $> -0.6$.
Provided that the plateau is truly primordial,
the constancy of the Li abundance in our plateau stars
indicates that our IRFM \teff scale is also 
highly reliable when we approach low metallicities,
and this would be consistent with the small dependence of the
IRFM on the detailed assumptions of model atmospheres.

Although the Li abundances derived in this work have
very small internal errors, 
systematic errors dominate the {\it absolute} abundances. 
For example, the use of Kurucz models with no
overshooting (NOVER, Castelli et al. 1997)
results in smaller $A_{Li}$ by about 0.08 dex. Note
that the Kurucz solar overshooting model reproduces
more solar observables than the NOVER model (Castelli et al. 1997).
Interestingly, the {\it (R-I)}$_{C}$ synthetic colors
computed from overshooting models
are in better agreement with the IRFM \teff scale 
than colors computed with NOVER models,
at least for solar metallicity dwarfs 
(6000 K $<$ \teff $<$ 6800 K, Mel\'endez \& Ram\'{\i}rez 2003)
and giants (4000 K $<$ \teff $<$ 6300 K, Ram\'{\i}rez \& Mel\'endez 2004a).

The mean Li abundance obtained for the plateau stars,
$A_{Li}$ = 2.37 dex, is considerably higher than the 
primordial Li abundance proposed by R01 ($A_{Li} \approx$ 2.0 - 2.1 dex).
As shown in Fig. 2, the reason for this
are the lower temperatures adopted by R01.
The temperature difference (between our \teff and those
adopted by R01) is metallicity dependent, 
and at very low metallicities ([Fe/H] $< -3)$ 
the \teff adopted by R01 is about 400 K lower than ours. 
When the temperatures adopted by R01 are used,
we also find a trend of Li abundance with metallicity.
If we use the same \teff (and $W_\lambda$) as those employed by
R01, $<A_{Li}>$ = 2.13 dex is found for stars with \teff $>$ 6000 K
and [Fe/H] $<$ -3, whereas R01 found $<A_{Li}>$ = 
2.07 dex for the same stars
(the small difference is due to the different model atmospheres adopted). 
Note that when Magain (1987) calibration is used in its validity 
range (\teff $<$ 6000 K, [Fe/H] $> -2$), the agreement 
with our \teff scale is satisfactory (open circles, Fig. 2);
outside that range it may not be appropriate to use it.

Most (0.07 dex) of the 0.13 dex difference between the present
Li plateau abundance and that obtained by BM97 is due to different
model atmospheres,
and the remaining difference is because our \teff scale is 
hotter for metal-poor F dwarfs.
We have compared our IRFM temperatures with those given by
Alonso et al. (1996) (employing the same reddening adopted
by Alonso et al.) The agreement is good for metal-poor stars 
with \teff $<$ 6500 K, and our temperatures are 
just $\approx$ 50 K higher in the range 6000 $<$ \teff $<$ 6500.
For hotter stars the differences are larger. 
Since only 5 stars with \teff $>$ 6500 K are included 
in the present work, their exclusion 
does not alter our results (Fig. 1), 
but restricts our claim of the constancy of the plateau 
to $-3 <$ [Fe/H] $< -1$ and 6000 K $<$ \teff $<$ 6500 K.

Ryan et al. (1999) have criticized Alonso's 
calibrations for turn-off stars
because of their strong metallicity dependence.
Nevertheless, this effect is predicted by both
Kurucz and MARCS model atmospheres. In fact, 
for metal-poor ([Fe/H] = -3) 
F dwarfs, our ($V-K$) IRFM calibration is in good agreement 
with $V-K$ colors predicted from MARCS models (Houdashelt
et al. 2000). At [Fe/H] = -3, our calibration (RM04c) gives 
6250 and 6500 K for $V-K_{2MASS}$ = 1.31 and 1.21, while Houdashelt
calibration results in \teff {\it higher} by +76 and +7 K,
respectively. The same exercise employing Kurucz models at $-2.5$
(M. S. Bessell 2004, private communication) gives a similar result.

Asplund et al. (2003) have shown that 3D hydrodynamical models with
NLTE computations give similar Li abundances than 1D + NLTE. 
However, that result was obtained neglecting collisions with H, 
and taking them into account, the Li abundance is
lowered (Barklem et al. 2003).
It is possible to compare 1D calculations 
(considering relative NLTE corrections) 
with 3D NLTE results by using the plateau star HD 84937, which has predicted
equivalent widths in 3D NLTE (Barklem et al. 2003). Employing
the same atmospheric parameters and Li abundance as given by 
Barklem et al. (2003), we obtain a 1D $W_\lambda$ = 11.4 m\AA,
which is very close to the 3D NLTE results, both without collisions
(11.1 m\AA) and when collisions are included (12.6 m\AA). Within
0.05 dex, both 3D NLTE calculations are in (probably fortuitous) 
agreement with the 1D calculation. 
Nevertheless, this result is by any means conclusive, 
as the 3D results are based on only one snapshot (Barklem et al. 2003).

The observational star-to-star scatter found in this work 
($\sigma_{obs}$ $\approx$ 0.05-0.06 dex) is considerably lower than those found 
in previous analyses of large samples: 0.13, 0.06-0.08, 0.089 and 0.094 dex 
according to Thorburn (1994), Spite et al. (1996), Ryan et al. (1996) and BM97, 
respectively. This extremely thin scatter of the plateau stars 
is fully explained by small errors in \teff and $W_\lambda$,
leaving little room (if any) 
for mixing or Galactic evolution.
Considering the constancy 
of the Li abundance of the plateau stars, depletion is
constrained probably at a level of 0.05 dex, and if
rotational mixing plays a role, then our mean $A_{Li}$ 
should be slightly increased.

Within the {\it absolute} abundance uncertainties involved (of the
order of 0.1 dex)
in our simple 1D modeling, 
we conclude that our results are just in marginal agreement with the 
constraint imposed by WMAP,
and if new physics or non-standard Big Bang nucleosynthesis
are invoked (e.g. Feng et al. 2003; Ichikawa \& Kawasaki 2004), 
the present work constraints their effects to a level of 0.1-0.15 dex.

\acknowledgements
We thank the referee for his constructive comments,
and JG Cohen for her suggestions. 
JM acknowledges support from NSF grant AST-0205951 to JGC.

\clearpage

\begin{figure}
\epsscale{}
\plotone{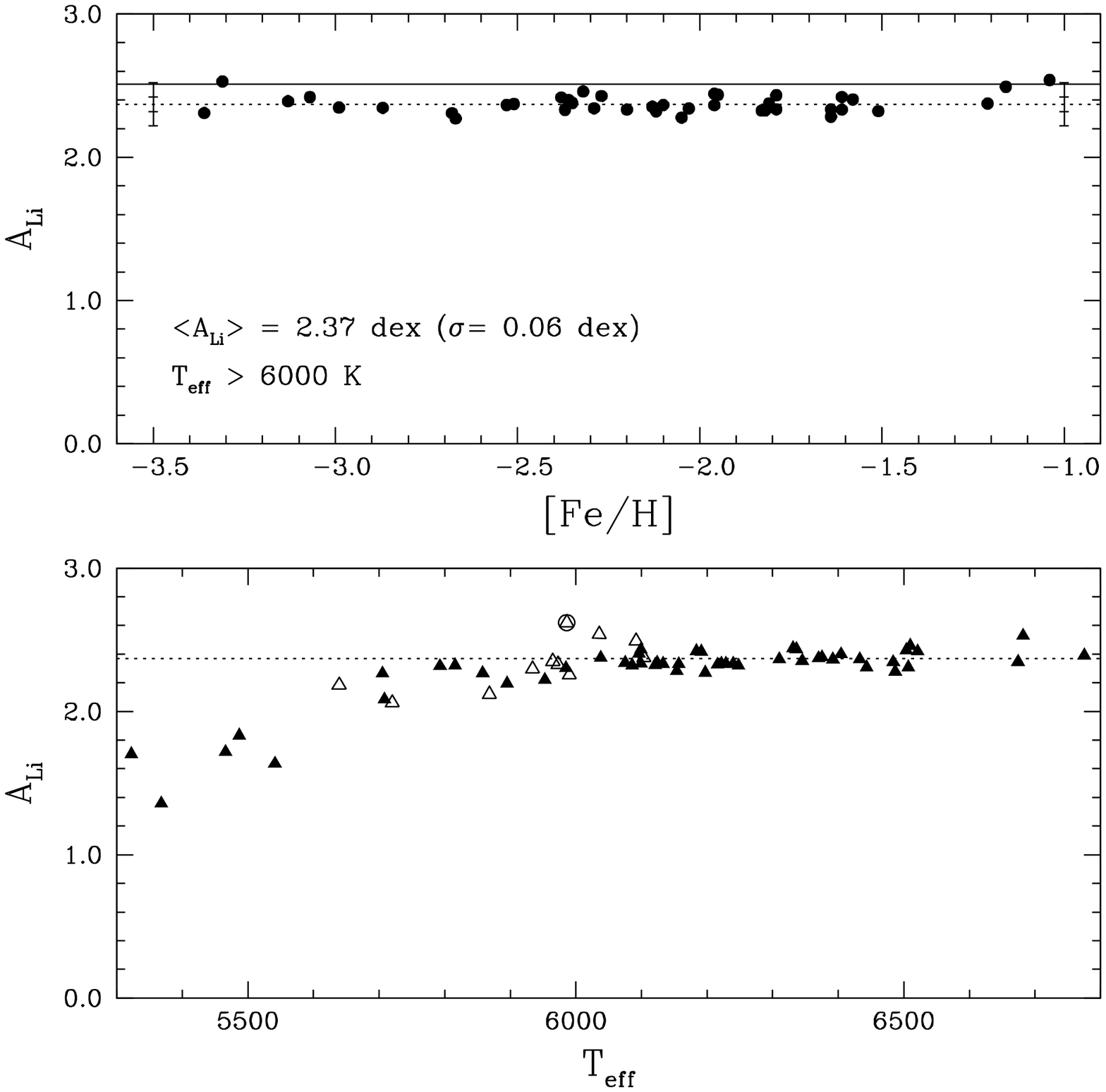}
\caption{Upper panel: 
$A_{Li}$ of the Spite plateau stars (\teff $>$ 6000 K)
as a function of [Fe/H]. 
Dotted lines indicate the mean
Li abundance of the plateau stars, and the solid line represents the
lower limit imposed by the WMAP constraint.
The error bars are the predicted error, 
$\sigma_{pred}$ (= 0.05 dex), and 3$\sigma_{pred}$ (= 0.15 dex).
Lower panel: $A_{Li}$ as a function of \teff
for stars with [Fe/H] $\leq -1.5$ (filled triangles) and 
[Fe/H] $> -1.5$ 
(open triangles). The star (HD 106038) with the highest 
Li abundance (open triangle inside the open circle) is a star
with peculiar abundances (Nissen \& Schuster 1997).
}
\end{figure}

\begin{figure}
\epsscale{}
\plotone{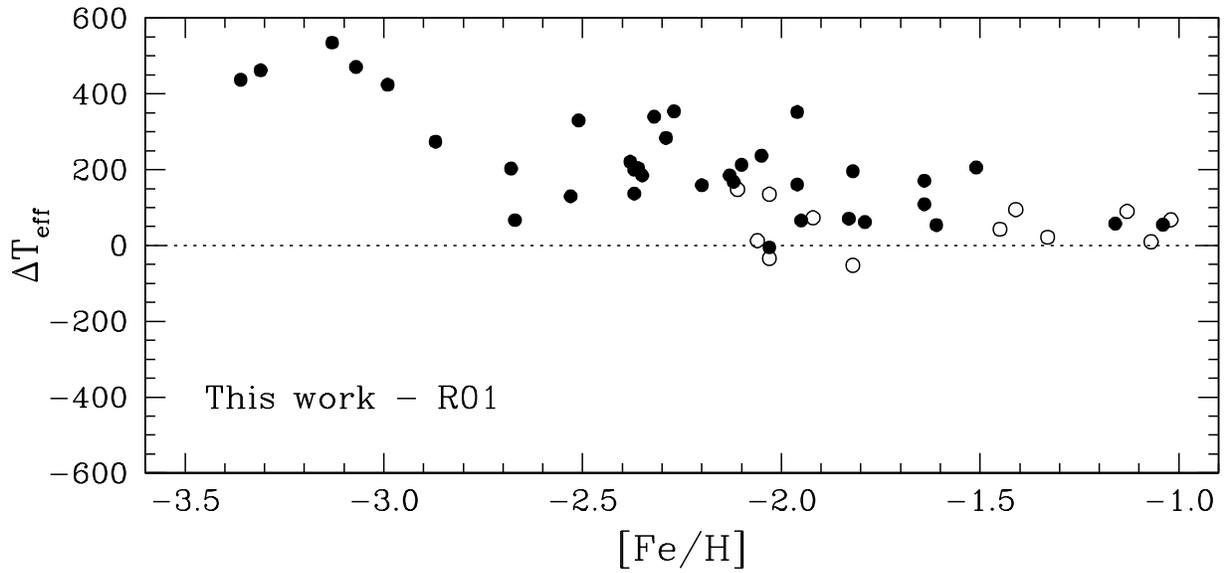}
\caption{Temperatures obtained in this work minus the temperatures 
from R01 (for stars in common with the present sample) 
as a function of the metallicities
adopted in the present work. Filled circles: plateau stars (\teff $>$ 6000 K);
open circles: stars with \teff $<$ 6000 K.}
\label{ryan}
\end{figure}

\begin{deluxetable}{lccccrcrrrrl}
\label{tab1}
\tablewidth{0pt}
\tabletypesize
\scriptsize
\tablecaption{Li abundances for the Spite plateau stars}
\tablehead{
\colhead{Star}  &
\colhead{${\rm T_{eff}}$} &
\colhead{log $g$} &
\colhead{[Fe/H]} & 
\colhead{$W_\lambda$} & 
\colhead{$A_{Li}$} \\
\colhead{}  &
\colhead{[K]} &
\colhead{[cm s$^{-2}$]} &
\colhead{[dex]} & 
\colhead{[m\AA]} & 
\colhead{[dex]}
}
\startdata
BD+002058   & 6092 &  4.19 & -1.16 & 41.9 & 2.49 \\
BD+023375   & 6124 &  3.97 & -2.29 & 32.5 & 2.34 \\
BD+030740   & 6443 &  3.76 & -2.68 & 19.5 & 2.31 \\
BD+090352   & 6075 &  4.38 & -2.03 & 34.0 & 2.34 \\
BD+092190   & 6487 &  4.11 & -2.05 & 17.3 & 2.28 \\
BD+174708   & 6154 &  3.93 & -1.64 & 25.8 & 2.28 \\
BD+203603   & 6248 &  4.15 & -2.12 & 26.2 & 2.32 \\
BD+241676   & 6510 &  3.91 & -2.32 & 25.3 & 2.46 \\
BD+262606   & 6157 &  4.20 & -2.37 & 30.5 & 2.33 \\
BD+282137   & 6229 &  3.50 & -2.20 & 27.2 & 2.33 \\
BD+342476   & 6433 &  3.92 & -2.10 & 22.6 & 2.37 \\
BD+422667   & 6086 &  4.13 & -1.51 & 30.6 & 2.32 \\
BD+592723   & 6121 &  4.00 & -1.83 & 30.5 & 2.33 \\
BD+710031   & 6331 &  4.05 & -1.96 & 31.0 & 2.44 \\
BD-043208   & 6404 &  3.77 & -2.36 & 25.4 & 2.40 \\
BD-100388   & 6240 &  3.61 & -2.37 & 27.0 & 2.33 \\
BD-133442   & 6484 &  3.98 & -2.87 & 20.5 & 2.35 \\
CD-2417504  & 6507 &  4.12 & -3.36 & 18.6 & 2.31 \\
CD-3018140  & 6336 &  4.08 & -1.95 & 30.3 & 2.44 \\
CD-3301173  & 6674 &  4.19 & -2.99 & 16.3 & 2.35 \\
CD-7101234  & 6375 &  4.43 & -2.35 & 26.0 & 2.38 \\
G059-024    & 6191 &  4.36 & -2.38 & 36.0 & 2.42 \\
G064-012    & 6682 &  4.10 & -3.31 & 23.4 & 2.53 \\
G064-037    & 6775 &  4.05 & -3.13 & 15.6 & 2.39 \\
G075-031    & 6036 &  4.07 & -1.04 & 48.0 & 2.54 \\
G192-043    & 6184 &  4.30 & -1.61 & 34.5 & 2.42 \\
G201-005    & 6370 &  3.79 & -2.51 & 25.0 & 2.37 \\
G206-034    & 6310 &  4.18 & -2.53 & 27.2 & 2.37 \\
HD016031    & 6216 &  4.07 & -1.82 & 27.0 & 2.33 \\
HD074000    & 6392 &  4.27 & -1.96 & 23.9 & 2.36 \\
HD084937    & 6345 &  3.96 & -2.13 & 24.8 & 2.35 \\
HD102200    & 6104 &  4.14 & -1.21 & 32.8 & 2.38 \\
HD108177    & 6099 &  4.30 & -1.64 & 31.4 & 2.33 \\
HD160617    & 6099 &  3.68 & -1.79 & 40.0 & 2.43 \\
HD166913    & 6096 &  4.08 & -1.58 & 37.0 & 2.40 \\
HD181743    & 6038 &  4.47 & -1.81 & 38.0 & 2.38 \\
HD218502    & 6222 &  3.94 & -1.79 & 27.1 & 2.34 \\
HD284248    & 6133 &  4.21 & -1.61 & 29.9 & 2.33 \\
HD338529    & 6504 &  3.90 & -2.27 & 23.9 & 2.43 \\
LP0056-0075 & 6197 &  4.36 & -2.67 & 25.5 & 2.27 \\
LP0831-0070 & 6521 &  4.23 & -3.07 & 23.8 & 2.42 \\
\enddata					  
\end{deluxetable}

\end{document}